\documentclass[11pt]{article}

\usepackage[preprint]{acl}
\usepackage{placeins}
\usepackage{times}
\usepackage{latexsym}
\usepackage[T1]{fontenc}

\usepackage[utf8]{inputenc}

\usepackage{microtype}

\usepackage{inconsolata}
\usepackage{hyperref}
\usepackage{enumitem}
\usepackage{graphicx}
\usepackage{adjustbox}
\usepackage{tikz}
\usepackage{pgfplots}
\pgfplotsset{compat=1.18}
\usepackage{multirow}
\usepackage{booktabs}
\usepackage{array}
\usepgfplotslibrary{groupplots}
\usepackage[ruled,vlined]{algorithm2e}

\usepackage{amsmath,amssymb}
\usepackage{subcaption}
\newcommand{\arxivincludegraphics}[2][]{%
  \IfFileExists{#2}{%
    \includegraphics[#1]{#2}%
  }{%
    \fbox{%
      \parbox[c][0.22\textheight][c]{0.95\linewidth}{%
        \centering
        Missing figure file:\\
        {\ttfamily\detokenize{#2}}%
      }%
    }%
  }%
}

\title{How Vulnerable Are Edge LLMs?}

\author{
{\small
Ao Ding\textsuperscript{1,\textdagger},
Hongzong Li\textsuperscript{2,\textdagger},
Zi Liang\textsuperscript{3,*},
Zhanpeng Shi\textsuperscript{4},
Shuxin Zhuang\textsuperscript{5,6},
Shiqin Tang\textsuperscript{6},
Rong Feng\textsuperscript{5,6},
Ping Lu\textsuperscript{7}
}\\
{\small
\textsuperscript{1}China University of Geoscience Beijing,
\textsuperscript{2}Hong Kong University of Science and Technology,
\textsuperscript{3}Hong Kong Polytechnic University,}\\
{\small
\textsuperscript{4}Jilin University,
\textsuperscript{5}City University of Hong Kong,
\textsuperscript{6}Chinese Academy of Sciences,
\textsuperscript{7}City University of Hong Kong (Dongguan)
}\\
{\small \texttt{aoding@email.cugb.edu.cn, lihongzong@ust.hk, zi1415926.liang@connect.polyu.hk, shizp9921@mails.jlu.edu.cn}}\\
{\small \texttt{shuxin.zhuang@my.cityu.edu.hk, shiqin.tang@cair.cas.org.hk, rongfeng3-c@my.cityu.edu.hk, 72405827@cityu-dg.edu.cn}}\\
{\small \textsuperscript{\textdagger}Equal contribution,\ \textsuperscript{*}Corresponding author}\\
{\small \textbf{Correspondence:} \href{mailto:zi1415926.liang@connect.polyu.hk}{zi1415926.liang@connect.polyu.hk}}
}

\begin{document}
\maketitle
\begin{abstract}
Large language models (LLMs) are increasingly deployed on edge devices under strict computation and quantization constraints, yet their security implications remain unclear. We study query-based knowledge extraction from quantized edge-deployed LLMs under realistic query budgets and show that, although quantization introduces noise, it does not remove the underlying semantic knowledge, allowing substantial behavioral recovery through carefully designed queries. To systematically analyze this risk, we propose \textbf{CLIQ} (\textbf{Cl}ustered \textbf{I}nstruction \textbf{Q}uerying), a structured query construction framework that improves semantic coverage while reducing redundancy. Experiments on quantized Qwen models (INT8/INT4) demonstrate that CLIQ consistently outperforms original queries across BERTScore, BLEU, and ROUGE, enabling more efficient extraction under limited budgets. These results indicate that quantization alone does not provide effective protection against query-based extraction, highlighting a previously underexplored security risk in edge-deployed LLMs.
\end{abstract}

\section{Introduction}

\begin{figure*}[t]
  \centering
  \arxivincludegraphics[width=\textwidth]{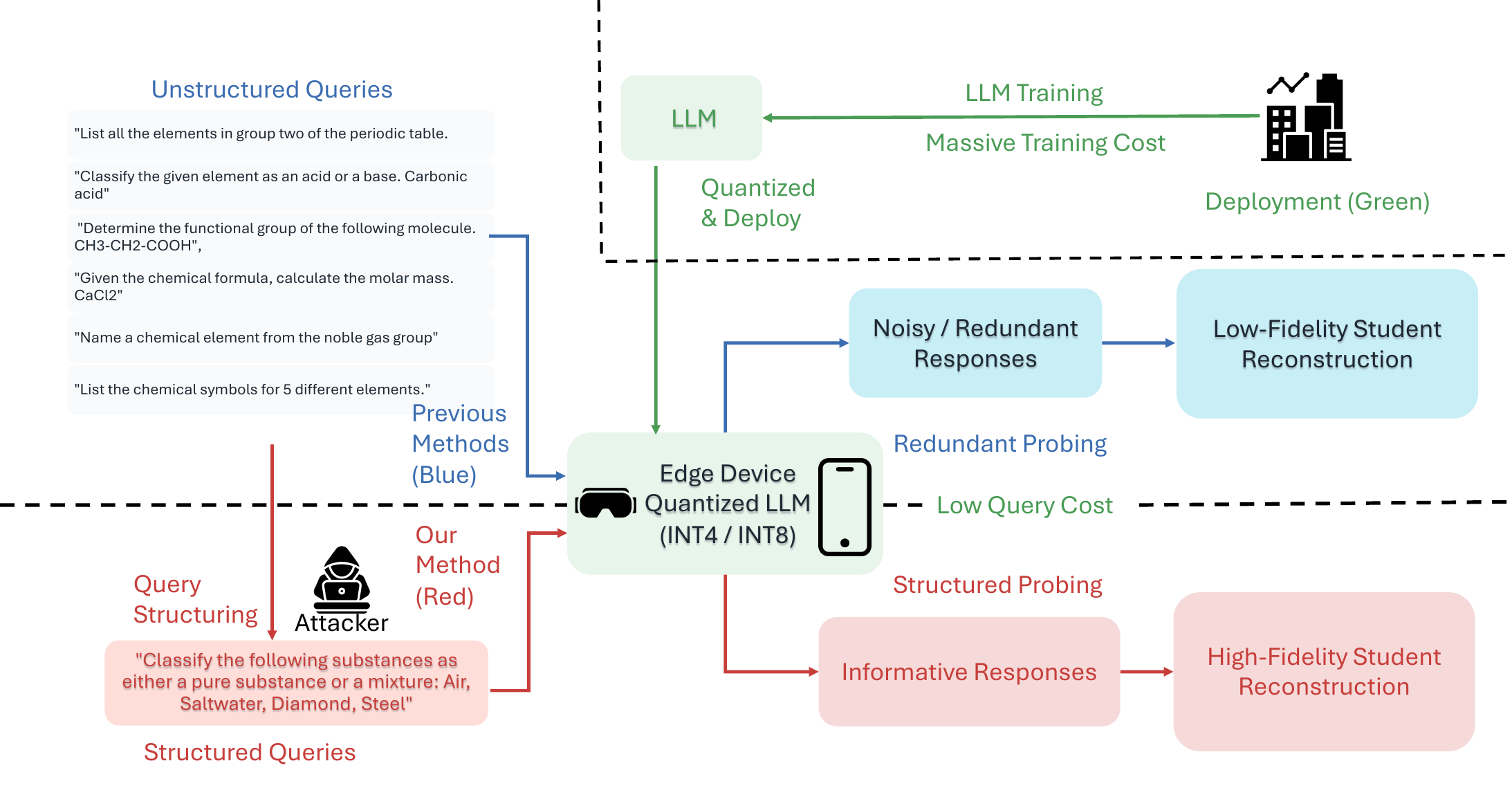}
\caption{
Overview of the proposed framework for query-based knowledge extraction
from edge-deployed quantized LLMs.
Previous approaches (blue) rely on unstructured queries, which often
lead to redundant probing and noisy responses, resulting in
low-fidelity reconstruction of model behavior.
CLIQ (red) performs structured query construction to produce more
informative responses under a limited query budget, enabling
high-fidelity student reconstruction of the edge model behavior.
}
  \label{fig:overview}
\end{figure*}

Large language models (LLMs) have achieved strong
instruction-following capabilities across a wide range of
applications \citep{DBLP:conf/nips/Ouyang}.
Training such models requires enormous computational
resources and financial investment, making trained LLMs
highly valuable intellectual assets \citep{hoffmann2022training}.
At the same time, the deployment landscape of LLMs is rapidly
shifting toward edge devices \citep{DBLP:journals/csur/ZhengCQSSC25}.
Recent industry developments reflect this trend:
Apple Intelligence integrates on-device foundation models
across iPhone, iPad, and Mac platforms
\citep{DBLP:journals/corr/abs-2507-13575},
Google deploys Gemini Nano on Android devices
\citep{google2023gemini},
Samsung Galaxy AI relies partly on local language models
\citep{samsung2024galaxyai}, and Qualcomm actively promotes
edge LLM deployment through mobile AI platforms
\citep{qualcomm2024genai}.
As LLMs increasingly operate directly on personal devices,
understanding the security implications of edge-deployed
models becomes increasingly important.

To enable such deployments, models are typically compressed
using aggressive low-bit quantization (e.g., INT8 or INT4),
which substantially reduces memory usage and inference
latency on resource-constrained devices.
However, quantization also introduces discretization noise
and reduces representational precision, leading to noisier
and less stable outputs
\citep{dettmers2023qlora,nagel2021quantization_whitepaper}.
This creates an important distinction from prior model
extraction and knowledge distillation settings, which mostly
focus on cloud-hosted or full-precision models and assume
that attackers can issue large numbers of queries to
approximate target behavior
\citep{DBLP:journals/corr/HintonVD15,TinyLlama}.
In contrast, edge-oriented scenarios impose much stricter
constraints: each interaction requires on-device inference,
and practical factors such as latency, energy consumption,
and interface limits naturally restrict the query budget
\citep{DBLP:journals/csur/ZhengCQSSC25}.
Under these conditions, naive random querying becomes highly
inefficient, as instruction queries sampled from natural
distributions often contain substantial semantic redundancy
and repeatedly probe similar capability regions
\citep{shu2026representation}.

From a security perspective, this raises a key question:
how much behavioral knowledge can still be recovered from
quantized edge models under realistic query budgets?
Although quantization may degrade the reliability of naive
extraction strategies
\citep{nagel2021quantization_whitepaper},
it does not remove the semantic knowledge encoded in model
parameters \citep{DBLP:journals/corr/abs-2103-13630}.
Instead, the effectiveness of query-based extraction depends
critically on how queries are constructed.
Motivated by this observation, we propose
\textbf{CLIQ} (\textbf{Cl}ustered \textbf{I}nstruction
\textbf{Q}uerying), a structured query probing framework
for analyzing the extractability of quantized edge LLMs.
As illustrated in Figure~\ref{fig:overview}, CLIQ organizes
candidate instruction queries through semantic clustering
and constructs representative queries for each cluster,
enabling broader coverage of the instruction space while
reducing redundant probing.

To quantify how much knowledge can be recovered through
query interactions, we collect responses from the edge
model and train a compact student model to approximate its
behavior.
If the student can reproduce the target model's responses
using only a limited set of query-response pairs, this
indicates that substantial behavioral knowledge has been
extracted through black-box interaction.
Extensive experiments across multiple model architectures,
quantization levels, and query budgets show that CLIQ
recovers significantly more model behavior than naive query
strategies.
Our results demonstrate that although quantization weakens
naive extraction methods, carefully structured queries
remain effective at recovering behavioral knowledge from
quantized edge models.

\noindent\textbf{Contributions.}
This paper makes the following contributions:
\begin{itemize}
\item[(i)]
We reveal a fundamental security risk in edge-deployed LLMs: quantization does not prevent query-based knowledge extraction, and structured querying enables efficient behavioral reconstruction under realistic query budgets. To the best of our knowledge, this is the first systematic study of extraction risks in quantized on-device LLMs.

\item[(ii)]
We propose CLIQ, a clustered instruction querying framework that constructs semantically structured queries to maximize coverage and minimize redundancy, enabling efficient probing of model capabilities under constrained budgets.

\item[(iii)]
We demonstrate across models and quantization levels that structured queries significantly improve extraction efficiency, showing that query design—not quantization—is the key factor governing model extractability in edge settings.
\end{itemize}

\section{Related Work}

\subsection{Knowledge Distillation for LLM}

Knowledge distillation transfers knowledge from large teacher models to smaller
students, with early work focusing on classification
\citep{DBLP:journals/corr/HintonVD15} and sequence modeling
\citep{DBLP:conf/emnlp/KimR16}, and recent studies extending to
instruction-level \citep{DBLP:conf/acl/HonovichSLS23} and
response-level distillation \citep{DBLP:conf/iclr/Gu0WH24}.
These approaches typically rely on large query sets and assume inexpensive,
high-quality teacher supervision.

Recent work explores data-free and API-based distillation, where students learn
from black-box teacher outputs without ground-truth annotations
\citep{DBLP:conf/iclr/ZhangCL23,chen2024knowledge}.
However, these methods treat black-box access as a practical constraint rather
than a security concern, implicitly assuming reliable supervision and ignoring
the risk of capability leakage through query interaction.

\subsection{Data-efficient and Query-efficient Distillation for LLM}

Data-efficient distillation methods reduce supervision cost via uncertainty-based
sampling \citep{du2025active} or synthetic data generation
\citep{DBLP:conf/acl/HonovichSLS23,DBLP:conf/acl/WangKMLSKH23}.
These approaches focus on selecting informative queries under the assumption
that the teacher model is high-precision and provides stable responses.

From a security perspective, this assumption does not hold in edge deployment
settings, where models are heavily quantized and query budgets are limited.
In such scenarios, query structure becomes a critical factor that governs both
distillation efficiency and the extent of extractable model behavior.

\subsection{Edge-oriented and Quantized LLMs}

Edge deployment has driven research on model compression and quantization,
including GPTQ and QLoRA
\citep{DBLP:journals/corr/abs-2210-17323,dettmers2023qlora},
which reduce memory and computation costs but introduce noise and reduced
expressivity.

Quantization is often assumed to mitigate extraction risks due to reduced
precision. However, its effect on query-based knowledge extraction remains
unclear. We address this gap by showing that structured queries can still
efficiently recover behavioral knowledge from quantized edge models.

\section{Method}

\begin{figure*}[t]
  \centering
  \arxivincludegraphics[width=\textwidth]{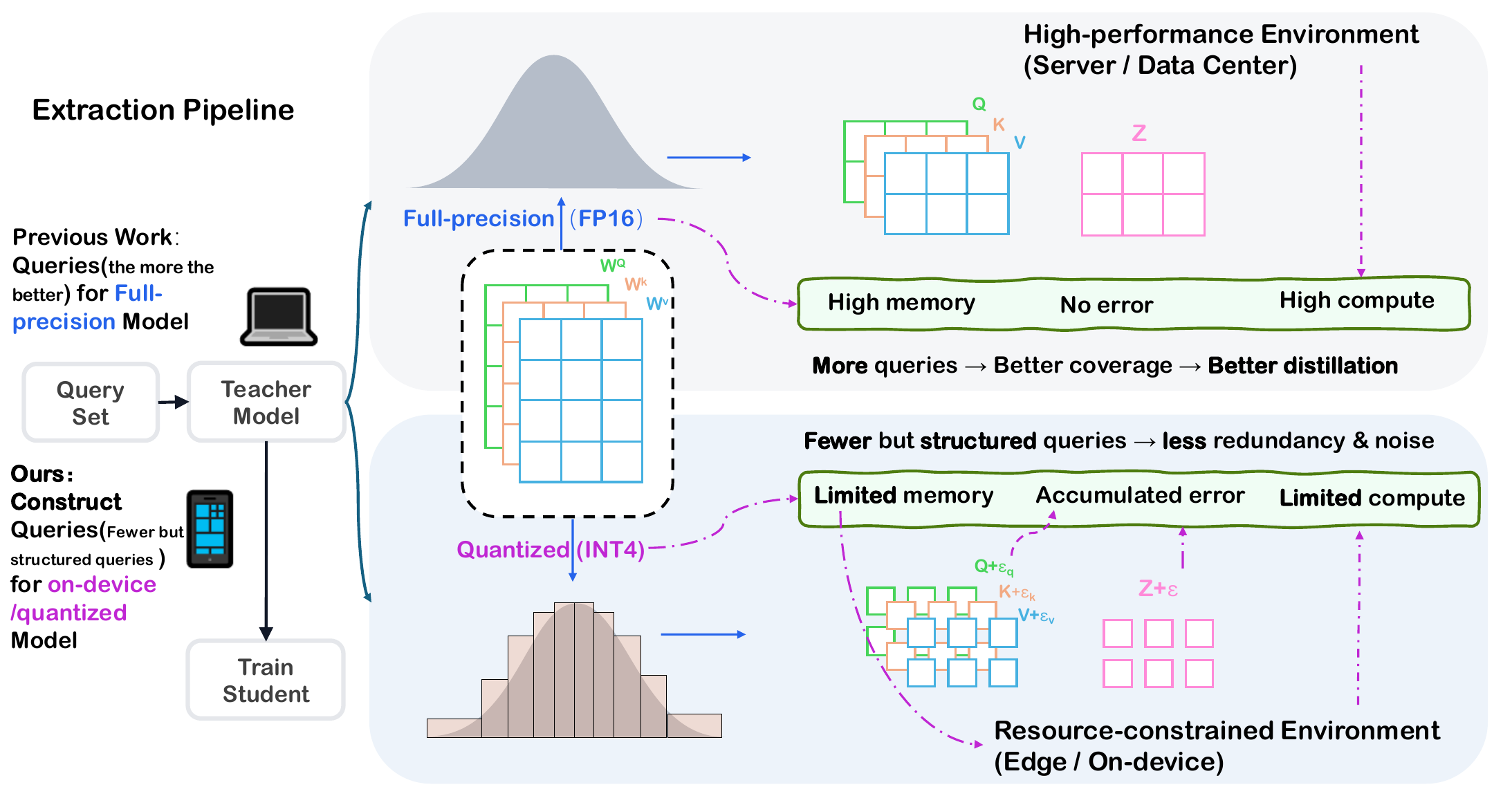}
    \caption{
        Threat framework for query-based knowledge extraction from
        quantized edge-deployed LLMs.
        Traditional extraction settings (top) assume full-precision
        teacher models in high-performance server environments,
        where abundant compute allows large-scale query probing.
        In contrast, edge-deployed LLMs (bottom) operate under
        quantization (e.g., INT4/INT8) and strict resource
        constraints, resulting in limited query budgets and noisy
        responses.
        Nevertheless, carefully structured queries can still
        recover substantial behavioral knowledge from the
        quantized model.
    }
  \label{fig:framework}
\end{figure*}

Figure~\ref{fig:framework} illustrates the threat framework
considered in this work.
Traditional model extraction studies typically assume
high-performance server environments and interact with
full-precision teacher models using a large number of
queries.
In contrast, our setting focuses on quantized LLMs deployed
on resource-constrained edge devices, where both query
budgets and computational resources are limited.
Our goal is to study whether carefully designed structured
queries can still recover substantial behavioral knowledge
from such models.

To systematically analyze this risk, we instantiate a
structured querying framework that organizes candidate
instructions according to semantic regions and constructs
representative probes for model interaction.
The collected responses are then used to reconstruct model
behavior through student modeling, providing a practical
measure of how much knowledge can be extracted from the
edge model under realistic deployment constraints.

\subsection{Threat Model and Problem Setting}

The increasing deployment of large language models (LLMs)
on edge devices such as mobile phones, personal assistants,
and embedded systems introduces a new security risk:
query-based interaction may expose behavioral knowledge
encoded in deployed models.
To satisfy memory and computation constraints, edge LLMs
are often heavily quantized (e.g., INT4 or INT8) and
exposed through query-based interfaces.

While such deployment enables efficient on-device
inference, it also creates an attack surface.
An adversary with query access can repeatedly interact
with the model and collect responses that reveal its
capabilities and decision boundaries.
As illustrated in Figure~\ref{fig:framework}, compared
with server-side extraction settings, the attacker
operates under a smaller query budget and a noisier
model interface, while the model still retains
substantial semantic knowledge.

\noindent\textbf{Adversary capability.}
The attacker has black-box access to the model and can
submit instruction queries and observe the generated
responses.
No access to model parameters, gradients, logits, or
training data is assumed.

\noindent\textbf{Query budget.}
The attacker is limited to a small number of queries due
to practical constraints such as rate limits, latency,
or device-side resource limits.

\noindent\textbf{Adversary goal.}
The goal is to recover as much information as possible
about the model's behavior using a limited query budget.
Following prior work on model extraction, we measure
this leakage by training a student model to imitate the
target responses.
High-fidelity imitation indicates that substantial
knowledge has been extracted through query interaction.

\subsection{Quantized Models as Leakage Surfaces}

Edge deployment commonly relies on aggressive weight
quantization to reduce model size and computational cost.
Although quantization reduces numerical precision and may
degrade the reliability of naive extraction strategies, it
does not remove the semantic knowledge encoded in model
parameters.
Instead, it transforms the extraction problem into a more
constrained and noisier variant.

We model the behavior of a quantized teacher model as
\begin{equation}
T_{\theta}^{\text{quant}}(x)
=
T_{\theta}^{\text{full}}(x) + \epsilon(x),
\end{equation}
where $\epsilon(x)$ denotes the output perturbation induced
by quantization.

Importantly, quantization should not be interpreted as a
security defense against query-based extraction.
Although the perturbation term may reduce the effectiveness
of naive random probing, the underlying behavioral and
semantic knowledge of the model remains largely preserved.
Therefore, an attacker who allocates queries more
efficiently may still recover substantial information from
the deployed model.
This motivates a structured query design strategy that is
robust to limited budgets and avoids excessive redundancy.

\subsection{CLIQ: Clustered Instruction Querying}

To study efficient knowledge extraction from quantized
edge models, we propose \textbf{CLIQ}
(\textbf{Cl}ustered \textbf{I}nstruction \textbf{Q}uerying),
a structured query construction framework for probing
model capabilities under limited query budgets.

Natural instruction queries exhibit strong semantic
redundancy, and naive sampling from a large instruction
pool often wastes queries on overlapping behaviors.
CLIQ addresses this issue by organizing queries into
semantic regions and generating representative probes
that maximize coverage while minimizing redundancy.
The framework consists of three stages: semantic query
clustering, cluster-aware query construction, and
behavior reconstruction.

\subsubsection{Semantic Query Partitioning}

Let the initial query pool be
\[
\mathcal{Q}_0 = \{q_1, q_2, \dots, q_N\}.
\]

Each query is mapped into a semantic embedding space
using a sentence encoder (e.g., Sentence-BERT).
We perform unsupervised clustering over the embeddings
to partition the pool into $K$ semantic clusters:
\[
\mathcal{Q}_0 = \bigcup_{k=1}^{K} \mathcal{Q}_k .
\]

Each cluster $\mathcal{Q}_k$ corresponds to a semantic
region in the instruction space.
This structured partition enables CLIQ to allocate
queries across distinct regions instead of relying on
redundant prompt sampling.

\subsubsection{Cluster-aware Query Construction}

Given the clusters, CLIQ generates a small set of
representative queries for each cluster.
For cluster $\mathcal{Q}_k$, we derive a cluster prototype
(e.g., centroid embeddings or representative seed queries)
and prompt a strong LLM to generate cluster-conditioned
instructions:
\[
\tilde{\mathcal{Q}}_k = \mathrm{LLM}(f(\mathcal{Q}_k)).
\]

The generated queries aim to ensure
(i) \emph{semantic representativeness},
(ii) \emph{intra-cluster diversity}, and
(iii) \emph{controlled complexity} to reduce instability
under quantization.

The final query set is
\[
\mathcal{Q} = \bigcup_{k=1}^{K} \tilde{\mathcal{Q}}_k .
\]

Compared with random sampling, this structured allocation
distributes the query budget across semantic regions and
improves probing efficiency under limited budgets.

\subsubsection{Behavior Reconstruction}

To measure information leakage through query interactions,
we train a student model $S_{\phi}$ on the query-response
pairs obtained from the teacher model.
Given the query set $\mathcal{Q}$ and teacher responses
$\{T_{\theta}(x)\}_{x \in \mathcal{Q}}$, the student
is trained by minimizing

\[
\mathcal{L}_{\text{KD}}
=
\frac{1}{|\mathcal{Q}|}
\sum_{x \in \mathcal{Q}}
\ell\bigl(S_{\phi}(x), T_{\theta}(x)\bigr),
\]

where $\ell(\cdot,\cdot)$ denotes the imitation loss.

If the student can reproduce the teacher's behavior
using only a limited set of structured queries, this
indicates that substantial knowledge has been extracted
through black-box interaction.

Overall, CLIQ shows that under a fixed query budget,
maximizing semantic coverage while reducing redundancy
significantly improves knowledge extraction efficiency
from quantized edge models.

\section{Experiments}
\label{sec:experiments}

\begin{table}[t]
\centering
\small
\resizebox{\linewidth}{!}{
\begin{tabular}{lcccc}
\toprule
\textbf{Model} & \textbf{BERT-F1} & \textbf{R1} & \textbf{RL} & \textbf{BLEU} \\
\midrule
Qwen2.5-7B (FP16)        & 84.51 & 20.93 & 14.87 & 2.81 \\
Qwen2.5-7B (GPTQ-INT4)  & 84.27 & 20.79 & 14.69 & 2.77 \\
\midrule
Qwen3-1.7B (FP16)       & 81.24 & 19.45 & 13.72 & 2.49 \\
Qwen3-1.7B (GPTQ-INT4)  & 80.98 & 19.27 & 13.56 & 2.43 \\
\bottomrule
\end{tabular}
}
\caption{Effect of GPTQ-based quantization on teacher model performance. Quantization introduces only marginal degradation across all metrics.All scores are reported in percentage (\%).}
\label{tab:teacher_quant}
\end{table}


\subsection{Experimental Setup}

\noindent\textbf{Teacher and Student Models.}
We use models from the Qwen family as both teachers and students.
Specifically, Qwen2.5-7B~\citep{qwen2025qwen25technicalreport} serves
as the teacher model, while edge-oriented students with up to 4B
parameters are used for extraction experiments.
To reflect realistic on-device deployment, all student models are
evaluated under aggressive quantization settings, including INT8
and INT4 using GPTQ.

Unless otherwise stated, both teacher inference and student training
use quantized weights, ensuring that the evaluation faithfully
reflects practical edge deployment constraints.

\noindent\textbf{Query Construction Strategies.}
We compare two query construction strategies under identical query
budgets:

\begin{itemize}[label={} ]

    \item[(i)] \textbf{Original Queries (OQ):} 
    instruction queries sampled from the original dataset.

    \item[(ii)] \textbf{CLIQ (Ours):} 
    cluster-aware generated instruction queries constructed via semantic clustering and cluster-aware prompting.

\end{itemize}

In all settings, training queries are randomly sampled from the
corresponding query pool during student training.
Thus, the only difference lies in how the query pool is constructed.

\noindent\textbf{Query Budget.}
Our main comparison uses a fixed query budget of 1000 queries.
We additionally analyze smaller budgets (100–400) and training
dynamics under a fixed budget of 500 queries.

\noindent\textbf{Evaluation Metrics.}
We report BERT-F1~\citep{bert_score}, BLEU~\citep{BLEU}, ROUGE-1,
ROUGE-2, ROUGE-L, and ROUGE-Lsum~\citep{rouge}.
All scores are averaged over held-out evaluation sets.

\subsection{Teacher Quantization Sanity Check}

Before evaluating extraction performance, we first examine the
effect of quantization on teacher models.
Table~\ref{tab:teacher_quant} compares full-precision and
GPTQ-quantized versions of Qwen2.5-7B and Qwen3-1.7B.

Across all metrics, GPTQ quantization introduces only marginal
performance degradation.
For example, Qwen2.5-7B-Instruct-GPTQ-Int4 achieves comparable
BERT-F1 and ROUGE scores to its FP16 counterpart.
These results confirm that quantized teachers remain sufficiently
expressive for query interaction and supervision.

\subsection{Main Results}
\label{sec:main_results}

\begin{table}[t]
\centering
\small
\resizebox{\linewidth}{!}{
\begin{tabular}{lcccccc}
\toprule
\textbf{Method} & \textbf{BERT-F1} & \textbf{BLEU}& \textbf{RLsum} \\
\midrule
Original Queries & 77.97 & 1.05 & 13.37 \\
\textbf{CLIQ (Ours)} & \textbf{84.35} & \textbf{2.77} & \textbf{17.50} \\
\bottomrule
\end{tabular}
}
\caption{Main distillation results on INT8-quantized 1.7B student models.All scores are reported in percentage (\%).}
\label{tab:main_results}
\end{table}


Table~\ref{tab:main_results} presents the main results on knowledge
extraction from quantized edge models.
Under identical query budgets, \textbf{CLIQ consistently outperforms
Original Queries} across all evaluation metrics.

Notably, a 1.7B student model with INT8 quantization distilled using
CLIQ achieves a BERT-F1 score of 0.8435, matching or even surpassing
the performance of significantly larger teacher models.
This result demonstrates that semantically representative queries can
effectively compensate for both limited model capacity and aggressive
quantization.

\subsection{Training Efficiency and Dynamics}

\begin{figure*}[t]
  \centering
  \arxivincludegraphics[width=\textwidth]{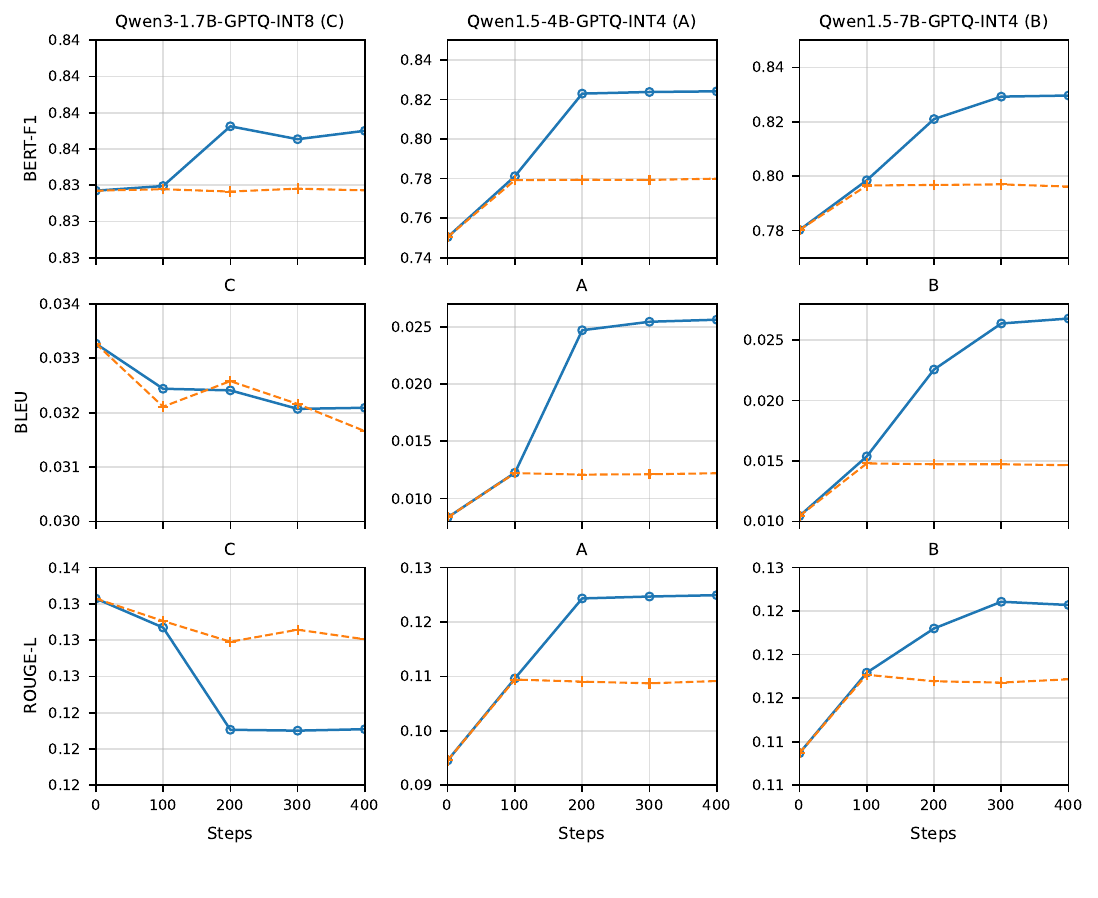}
\caption{
Effect of training steps under a fixed query budget (500 queries) across three student models.
We report BERT-F1, BLEU, and ROUGE-L, and include the 0-step base model as initialization.
CLIQ (ours) consistently outperforms OQ and exhibits diminishing returns after $\sim$200--300 steps.
}
\label{fig:steps_ablation_allmetrics}
\end{figure*}

Figure~\ref{fig:steps_ablation_allmetrics} shows training
dynamics under a fixed query budget of 500 queries.
CLIQ consistently outperforms Original Queries (OQ).

For BERT-F1, CLIQ achieves rapid improvements within the
first 200 steps and saturates around 200–300 steps,
while OQ-trained models exhibit nearly flat trajectories.

\begin{table}[t]
\centering
\small
\setlength{\tabcolsep}{4pt}
\renewcommand{\arraystretch}{1.1}
\begin{tabular}{c c c c c c}
\toprule
Quant. & Query & Steps & BERT-F1 & BLEU & ROUGE-L \\
\midrule

\multirow{6}{*}{INT4}
& Original & 100 & 80.32 & 1.55 & 13.83 \\
&          & 300 & 80.44 & 1.49 & 14.12 \\
&          & 500 & 80.03 & 1.42 & 13.41 \\
& CLIQ     & 100 & 78.20 & 1.08 & 12.45 \\
&          & 300 & 82.86 & 2.56 & 14.22 \\
&          & 500 & \textbf{82.92} & \textbf{2.55} & \textbf{14.54} \\

\midrule

\multirow{6}{*}{INT8}
& Original & 100 & 80.14 & 1.50 & 13.49 \\
&          & 300 & 80.17 & 1.41 & 13.51 \\
&          & 500 & 80.08 & 1.43 & 13.49 \\
& CLIQ     & 100 & 78.18 & 1.07 & 12.53 \\
&          & 300 & 82.70 & 2.44 & 14.48 \\
&          & 500 & 82.74 & 2.47 & 14.51 \\

\bottomrule
\end{tabular}
\caption{
Effect of training steps under different quantization settings.
CLIQ shows consistent improvements from 100 to 300 steps,
while Original queries exhibit minimal gains.
}
\label{tab:training_steps}
\end{table}

Table~\ref{tab:training_steps} further analyzes training
dynamics under different quantization settings.
Across both INT4 and INT8, two consistent patterns emerge:
(i) OQ shows minimal gains across training steps, and
(ii) CLIQ achieves rapid improvement from 100 to 300 steps,
followed by saturation.

CLIQ starts slightly lower at early steps but quickly
surpasses OQ, suggesting that broader semantic coverage
requires initial optimization but leads to stronger
final performance.

\subsection{Query Budget and Sample Efficiency}

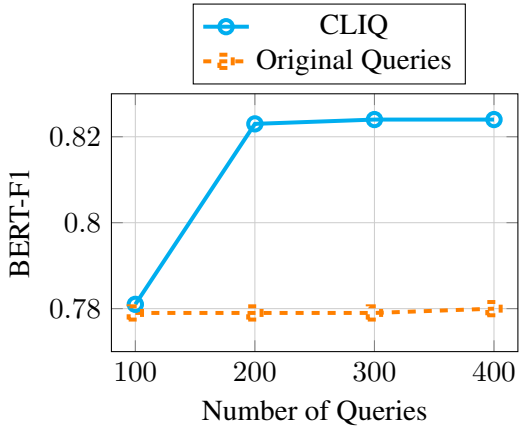
\begin{figure}[t]
\centering
\begin{tikzpicture}
\begin{axis}[
    width=0.9\linewidth,
    height=5cm,
    xlabel={Number of Queries},
    ylabel={BERT-F1},
    xmin=80, xmax=420,
    ymin=0.77, ymax=0.83,
    xtick={100,200,300,400},
    ytick={0.78,0.80,0.82},
    legend style={at={(0.2,1.35)}, anchor=north west},
    grid=both,
    grid style={line width=.1pt, draw=gray!20},
    major grid style={line width=.2pt,draw=gray!40}
]

\addplot[
    color= cyan,
    mark=o,
    line width=1.5pt,
    mark size=2.5pt
]
coordinates {
    (100,0.781)
    (200,0.823)
    (300,0.824)
    (400,0.824)
};
\addlegendentry{CLIQ}

\addplot[
    color=orange,
    mark=square,
    dashed,
    line width=1.5pt,
    mark size=2.5pt
]
coordinates {
    (100,0.779)
    (200,0.779)
    (300,0.779)
    (400,0.780)
};
\addlegendentry{Original Queries}

\end{axis}
\end{tikzpicture}
\caption{Effect of query budget on distillation performance.
CLIQ exhibits rapid gains as the number of queries increases from 100 to 300, followed by diminishing returns, while Original Queries show minimal improvement.}
\label{fig:query_budget}
\end{figure}


We further analyze sample efficiency by varying the number of
generated queries from 100 to 400.
As shown in Figure~\ref{fig:query_budget}, CLIQ improves rapidly
as the query budget increases from 100 to 300 and then gradually
saturates.

In contrast, Original Queries improve more slowly and reach a
performance plateau earlier, highlighting the superior efficiency
of structured query construction.

\subsection{Effect of Quantization}

\begin{table}[t]
\centering

\small
\setlength{\tabcolsep}{6pt}
\renewcommand{\arraystretch}{1.2}
\begin{tabular}{lcc}
\toprule
\textbf{Query Strategy} & \textbf{INT8} & \textbf{INT4} \\
\midrule
OQ & 80.08 & 80.03 \\
\textbf{CLIQ (ours)} & \textbf{82.74} & \textbf{82.92} \\
\bottomrule
\end{tabular}
\vspace{0.5em}
\begin{flushleft}
\footnotesize
\textbf{Note:} BERT-F1 scores are reported.
All results are obtained with a fixed query budget (500 queries) and 500 training steps. All scores are reported in percentage (\%).
\end{flushleft}
\caption{Effect of quantization on different query construction strategies
using a 1.8B edge-oriented student model.}
\label{tab:quantization_effect}
\end{table}


Finally, we evaluate extraction performance under different
quantization levels.
Across both INT8 and INT4 settings, CLIQ consistently outperforms
Original Queries.

Under INT4 quantization, CLIQ achieves a BERT-F1 score of 0.829
compared to 0.800 obtained by Original Queries.
These results show that structured query construction remains
effective even under aggressive low-precision deployment.

\section{Analysis and Discussion}

In this section, we analyze why cluster-aware query construction
is particularly effective for distilling quantized edge-oriented
language models and discuss the limitations of the proposed
approach.

\subsection{Why Does Query Clustering Help Quantized Students?}

Quantized student models have significantly reduced effective
capacity due to both low-precision weights (e.g., INT8 or INT4)
and smaller parameter counts.
As a result, they are more sensitive to noisy, redundant, or
poorly structured supervision signals.

Query clustering explicitly models the latent structure of the
instruction space by grouping semantically similar queries.
This allows CLIQ to reduce redundancy in the query pool and
encourage coverage across distinct semantic regions.
For quantized students, such structured supervision is
particularly beneficial: instead of allocating limited capacity
to highly correlated instruction patterns, the model can focus
on learning representative behaviors across different semantic
regions.

It is important to note that in all experimental settings,
training queries are randomly sampled during optimization;
the only difference lies in the structure of the query pool.
The observed improvements therefore arise purely from the
semantic organization of queries rather than from training
procedures.
This effect becomes more pronounced under aggressive
quantization (e.g., INT4), where structured supervision helps
mitigate the impact of quantization noise.

\subsection{Effect of Cluster-aware Query Generation}

Beyond clustering itself, the cluster-aware query generation
process plays an important role in improving distillation
quality.
Unlike generic instruction generation, our prompts are
conditioned on cluster-level semantics and explicitly constrain
instruction complexity, length, and reasoning depth.

These constraints are particularly important for edge-oriented
models, which often struggle with long reasoning chains or
overly complex instructions.
By generating queries that are both semantically representative
and appropriately scoped, CLIQ produces supervision signals
that better match the inductive biases and capacity limits of
small quantized transformer models.

Empirically, this design choice leads to consistent improvements
in ROUGE-based metrics, suggesting that the gains arise from
better coverage of salient semantic content rather than
superficial fluency improvements.

\subsection{Sample Efficiency and Performance Saturation}

Our experiments show that student performance improves rapidly
as the number of cluster-aware generated queries increases from
100 to approximately 300, after which the gains gradually
saturate.
This behavior suggests that cluster-aware query generation is
highly sample-efficient: a relatively small number of
well-structured queries is sufficient to capture most
informative instruction patterns.

In contrast, models trained on original query pools improve
more slowly and reach performance saturation earlier.
This indicates that simply increasing the number of queries is
insufficient for efficient knowledge extraction; the semantic
quality and diversity of queries play a more critical role
under strict query budgets.


\section{Conclusion}

In this work, we study the security implications of deploying
quantized large language models on edge devices.
While quantization is commonly used to enable efficient
on-device inference, our results show that it does not prevent
knowledge leakage through query-based interactions.
Instead, carefully designed queries can still extract
substantial behavioral knowledge from quantized models
under limited query budgets.

To systematically analyze this risk, we propose
\textbf{CLIQ}, a cluster-aware query construction framework
that organizes instruction queries according to their
semantic structure and generates representative probes for
efficient model interaction.
Extensive experiments demonstrate that CLIQ significantly
improves extraction efficiency compared with naive query
sampling, particularly for heavily quantized student models.

These findings suggest that structured query design plays a
critical role in knowledge extraction from edge-deployed
LLMs and highlight the need for stronger safeguards when
deploying quantized language models on resource-constrained
devices.
Future work may explore defenses against structured
query-based extraction and extend our analysis to broader
edge deployment scenarios.

\section{Limitations}

While our results demonstrate the effectiveness of structured
query design, several limitations remain.

\paragraph{Limited Query Budget Regime.}
Our method assumes moderate query budgets (e.g., 100–1000).
Under extremely small budgets, both CLIQ and original queries
perform poorly, as insufficient interaction limits recoverable
behavior (see Figure~\ref{fig:query_budget}).

\paragraph{Dependence on Query Coverage.}
CLIQ improves semantic coverage but does not guarantee it.
If certain capability regions are absent from the initial
query pool, they may remain unexplored despite clustering,
as reflected by incomplete cluster coverage in low-budget
settings (Figure~\ref{fig:cluster_hit_rate}).

\paragraph{Sensitivity to Quantization Noise.}
Although CLIQ mitigates quantization noise, extreme
low-precision settings still degrade extraction quality.
For instance, INT4 models consistently yield lower scores
than INT8 across all settings (Table~\ref{tab:quantization_effect}),
indicating a fundamental limit imposed by noisy responses.

\paragraph{Scope of Evaluation.}
Our experiments focus on instruction-following benchmarks.
It remains unclear how structured querying performs in more
complex settings such as tool-augmented agents or models
with strong safety filtering.
\section{Ethical Considerations}

This work studies query-efficient knowledge extraction from
edge-deployed and quantized LLMs, which raises potential
concerns regarding model misuse and unauthorized behavioral
transfer.

\paragraph{Model Extraction Risks.}
Our results show that quantization does not eliminate the
risk of knowledge extraction under query-based interaction.
Although CLIQ does not recover model parameters or exact
output distributions, it improves the efficiency of
behavioral imitation under limited query budgets.
This highlights a practical security risk in edge deployment,
rather than a vulnerability specific to our method.

\paragraph{Responsible Use and Compliance.}
All experiments are conducted using publicly available data
and standard, rate-limited API access.
We do not bypass safeguards or violate usage policies.
We emphasize that applying such techniques should comply
with model licenses, terms of service, and local regulations.

\paragraph{Broader Impact.}
Our goal is to better understand the security implications
of deploying LLMs on resource-constrained devices.
By identifying how structured queries affect extractability,
this work can inform the design of more robust deployment
strategies and defenses against misuse.




\appendix

\clearpage
\section{Appendix}

\subsection{CLIQ Pipeline Details}
\label{app:cliq_pipeline}

\subsubsection{Query Representation and Clustering}
\label{app:clustering}

\paragraph{Query Pool Construction.}
Given a raw instruction dataset in JSON format, we construct the text query for
each sample by concatenating the \texttt{instruction} field and the optional
\texttt{input} field:
\[
q = \texttt{instruction} \oplus \texttt{input}.
\]
If \texttt{input} is empty, we use \texttt{instruction} only.

\paragraph{Sentence Embeddings.}
Each query is encoded using a sentence encoder
(\texttt{sentence-transformers/all-MiniLM-L6-v2}).
We apply mean pooling with attention masking followed by $\ell_2$ normalization:
\[
\mathbf{e}(q)=\text{Normalize}\big(\text{MeanPool}(\text{LM}(q))\big).
\]
The maximum sequence length is set to 512 with a batch size of 64.

\paragraph{Clustering.}
All embeddings are clustered using MiniBatchKMeans with $K=100$ clusters and a
fixed random seed (42).
MiniBatchKMeans \citep{kmeans} is adopted for scalability and memory efficiency.

\paragraph{Filtering Small Clusters.}
Clusters whose size falls below a threshold (default: 5) are removed to avoid
unstable prototypes and overly sparse supervision.
Remaining clusters are re-indexed to ensure contiguous cluster IDs.

\subsubsection{Cluster-aware Query Generation}
\label{app:generation}

\paragraph{Cluster-conditioned Prompting.}
For each cluster, we construct a prompt containing up to $M$ in-cluster examples
(default: $M=1000$) to characterize the cluster semantics and instruction style.
In our main experiments, we generate $m=10$ queries per cluster unless otherwise stated,
while for visualization we report a more query-efficient setting with $m=5$.

\paragraph{Teacher API and Decoding.}
We use a Qwen-family teacher accessed via an OpenAI-compatible API endpoint.
The decoding temperature is set to 0.7 with a large token budget
(max tokens: 16384) to reduce truncation.

\paragraph{Robust Output Parsing.}
The teacher is instructed to output a JSON array of
\{\texttt{instruction}, \texttt{input}\} pairs.
We apply a robust parser that extracts JSON content from markdown blocks and
repairs truncated outputs by retaining the last complete object and closing the
array.

\subsubsection{Implementation Details and Hyperparameters}
\label{app:hyperparams}

Table~\ref{tab:hyperparams} summarizes the key hyperparameters used in clustering
and query generation.
All settings are fixed across experiments unless otherwise stated.

\begin{table}[t]
\centering
\small
\setlength{\tabcolsep}{5pt}
\renewcommand{\arraystretch}{1.15}
\resizebox{\columnwidth}{!}{
\begin{tabular}{l l}
\toprule
\textbf{Component} & \textbf{Setting} \\
\midrule
Sentence encoder & \texttt{sentence-transformers/all-MiniLM-L6-v2} \\
Max length & 512 \\
Embedding batch size & 64 \\
Pooling & mean pooling w/ attention mask \\
Normalization & $\ell_2$ normalization \\
Clustering algorithm & MiniBatchKMeans \\
clusters $K$ & 100 \\
Seed & 42 \\
MiniBatch size & $\min(1000, |\mathcal{Q}|/10)$ \\
Cluster size filter & min size = 5 \\
\midrule
Teacher API model & \texttt{Qwen/Qwen3-30B-A3B-Instruct-2507} \\
Temperature & 0.7 \\
Max tokens & 16384 \\
Timeout & 300s \\
Retries & 5 (exponential backoff) \\
Examples per cluster $M$ & 1000 \\
Generated queries/cluster $m$ & 10 \\
\bottomrule
\end{tabular}
}
\caption{Hyperparameters for clustering and cluster-aware query generation.}
\label{tab:hyperparams}
\end{table}


\begin{algorithm}[t]
\caption{CLIQ Pipeline}
\label{alg:cliq_pipeline}
\small
\KwIn{Query pool $\mathcal{Q}$; clusters $K$; min size $s$; examples $M$; queries $m$}
\KwOut{Generated query set $\widetilde{\mathcal{Q}}$}

Encode each query $q \in \mathcal{Q}$ into $\mathbf{e}(q)$\;
Cluster embeddings into $K$ clusters via MiniBatchKMeans\;
Remove clusters with size $< s$ and re-index\;

$\widetilde{\mathcal{Q}} \leftarrow \emptyset$\;
\ForEach{cluster $c$}{
  Sample up to $M$ in-cluster queries\;
  Construct cluster-conditioned prompt\;
  Query teacher to generate $m$ new queries\;
  Parse and repair JSON outputs\;
  $\widetilde{\mathcal{Q}} \leftarrow \widetilde{\mathcal{Q}} \cup$ generated queries\;
}
\Return{$\widetilde{\mathcal{Q}}$}\;
\end{algorithm}

\FloatBarrier

\subsection{Additional Quantitative Results}
\label{app:quant_results}

\subsubsection{Full Model Evaluation}
\label{app:full_results}

Table~\ref{tab:model_results} reports the full evaluation results under a unified
teacher--student setup, where \textbf{Qwen2.5-7B} serves as the teacher and
\textbf{Qwen3-1.7B} is the student architecture.
We report (i) the teacher under FP16 and Int4, (ii) base (unaligned) students
under FP16 and Int8, and (iii) aligned students trained with either original
queries (OQ) or cluster-aware generated queries (CLIQ), including validation
results (``valid'').

\paragraph{Teacher vs.\ Base Student.}
The base student is substantially weaker than the teacher across all metrics
(e.g., BERT-F1 and ROUGE family), which motivates alignment/distillation.
Quantization of the base student (FP16 vs.\ INT8) changes performance only
slightly, indicating that the base capability gap is primarily due to model size
rather than precision.

\paragraph{Aligned Students with OQ vs.\ CLIQ.}
Alignment using queries consistently improves the student over its base version.
Comparing query strategies, CLIQ achieves the strongest overall results among
student settings, improving BERT-F1, BLEU, and ROUGE metrics over OQ.
This pattern supports the central claim that \emph{query quality and semantic
coverage}, rather than merely the existence of alignment, is a key driver of the
observed gains.

\paragraph{Validation Consistency.}
For both OQ and CLIQ, ``valid'' results are close to the corresponding test
results, suggesting that the improvements are not due to overfitting to the test
set but reflect stable generalization.


\begin{table*}[t]
\centering
\small
\setlength{\tabcolsep}{6pt}
\renewcommand{\arraystretch}{1.15}
\begin{adjustbox}{max width=\textwidth}
\begin{tabular}{llclcccccc}
\toprule
\textbf{Model} & \textbf{Role} & \textbf{Quant.} & \textbf{Query} &
\textbf{BERT-F1} & \textbf{BLEU} & \textbf{R-1} & \textbf{R-2} & \textbf{R-L} & \textbf{R-Lsum} \\
\midrule
\multirow{2}{*}{Qwen2.5-7B} & \multirow{2}{*}{Teacher} & INT4 & -- 
& 84.22 & \textbf{3.15 }& 21.92 & 9.58 & 16.56 & 18.63 \\
                           &                          & FP16 & -- 
& 83.69 & 2.94 & 22.08 & 9.36 & 16.85 & 18.76 \\
\midrule

\multirow{6}{*}{Qwen3-1.7B} & \multirow{2}{*}{Base}    & INT8 & -- 
& 81.40 & 1.73 & 12.23 & 4.36 & 8.70 & 10.03 \\
                           &                          & FP16 & -- 
& 81.53 & 1.70 & 12.35 & 4.46 & 8.92 & 10.16 \\
                           
\cmidrule{2-10}
& \multirow{4}{*}{Student} 
& \multirow{2}{*}{FP16$\rightarrow$INT8} & OQ        
& 83.75 & 2.50 & 22.30 & 9.49 & 16.68 & 18.94 \\
                           &                          &                                      & OQ (valid) 
& 83.78 & 2.46 & 22.24 & 9.57 & 16.83 & 18.96 \\
\cmidrule{3-10}
                           &                          & \multirow{2}{*}{INT4$\rightarrow$INT8} & CLIQ        
& \textbf{84.35} & 3.04 & \textbf{23.42} & \textbf{10.60} & \textbf{17.60} & \textbf{19.87} \\
                           &                          &                                       & CLIQ (valid) 
& 84.23 & 2.92 & 23.07 & 10.26 & 17.27 & 19.53 \\
\bottomrule
\end{tabular}
\end{adjustbox}
\caption{Comprehensive Evaluation Results under Different Model, Quantization, and Query Settings.
All scores are reported in percentage (\%).}
\label{tab:model_results}
\end{table*}


\subsubsection{Training Dynamics Analysis}
\label{app:training_dynamics}

\paragraph{Fixed Query Budget.}
Figure~\ref{fig:steps_ablation_allmetrics} analyzes the training dynamics under a fixed query
budget of 500 queries across three edge-oriented student models, reporting
BERT-F1, BLEU, and ROUGE-L, with the 0-step base model included as initialization.

Across all models and metrics, CLIQ exhibits clear and consistent advantages over
original queries (OQ).
For BERT-F1, CLIQ leads to rapid performance gains within the first 200 training
steps for all three students, followed by saturation around 200--300 steps,
indicating diminishing returns once the structured supervision signal has been
fully exploited.
In contrast, OQ-trained models remain largely insensitive to additional training
steps, showing near-flat trajectories across all models.

The same qualitative trend is observed for BLEU and ROUGE-L, where CLIQ yields
substantial improvements over OQ, particularly for the smaller and more heavily
quantized student models (A and B).
Notably, under OQ supervision, BLEU and ROUGE-L remain almost unchanged throughout
training, whereas CLIQ enables sustained improvements as optimization proceeds,
highlighting its stronger and less redundant supervision signal.

Including the 0-step initialization further reveals that CLIQ consistently
outperforms OQ beyond early training, rather than merely accelerating convergence.
Together, these results demonstrate that under a fixed query budget, performance
gains primarily arise from improved query quality and semantic coverage, rather
than from increased optimization alone, and that CLIQ fundamentally alters the
learning dynamics of quantized student models.

%
\begin{figure*}[t]
\centering
\setlength{\tabcolsep}{6pt}

\begin{subfigure}[t]{0.32\textwidth}
\centering
\begin{tikzpicture}
\begin{axis}[
    width=\linewidth, height=4.6cm,
    xlabel={Steps}, ylabel={BERT-F1},
    xtick={100,200,300,400},
    ymin=0.77, ymax=0.85,
    grid=both,
    tick label style={font=\small},
    label style={font=\small},
    legend style={font=\small, at={(0.5,1.15)}, anchor=south, legend columns=2},
]
\addplot[mark=o, thick] coordinates {(100,0.83494) (200,0.83906) (300,0.83817) (400,0.83875)};
\addlegendentry{CLIQ(ours)}
\addplot[mark=square, thick, dashed] coordinates {(100,0.83472) (200,0.83456) (300,0.83476) (400,0.83464)};
\addlegendentry{OQ}
\end{axis}
\end{tikzpicture}
\caption{Qwen3-1.7B-GPTQ-Int8}
\end{subfigure}
\hfill
\begin{subfigure}[t]{0.32\textwidth}
\centering
\begin{tikzpicture}
\begin{axis}[
    width=\linewidth, height=4.6cm,
    xlabel={Steps},
    xtick={100,200,300,400},
    ymin=0.77, ymax=0.85,
    grid=both,
    tick label style={font=\small},
    label style={font=\small},
    legend style={font=\small, at={(0.5,1.15)}, anchor=south, legend columns=2},
]
\addplot[mark=o, thick] coordinates {(100,0.78119) (200,0.82297) (300,0.82379) (400,0.82413)};
\addlegendentry{CLIQ(ours)}
\addplot[mark=square, thick, dashed] coordinates {(100,0.77921) (200,0.77941) (300,0.77932) (400,0.77997)};
\addlegendentry{OQ}
\end{axis}
\end{tikzpicture}
\caption{Qwen1.5-4B-GPTQ-Int4}
\end{subfigure}
\hfill
\begin{subfigure}[t]{0.32\textwidth}
\centering
\begin{tikzpicture}
\begin{axis}[
    width=\linewidth, height=4.6cm,
    xlabel={Steps},
    xtick={100,200,300,400},
    ymin=0.77, ymax=0.85,
    grid=both,
    tick label style={font=\small},
    label style={font=\small},
    legend style={font=\small, at={(0.5,1.15)}, anchor=south, legend columns=2},
]
\addplot[mark=o, thick] coordinates {(100,0.79846) (200,0.82098) (300,0.82923) (400,0.82964)};
\addlegendentry{CLIQ(ours)}
\addplot[mark=square, thick, dashed] coordinates {(100,0.79656) (200,0.79675) (300,0.79695) (400,0.79616)};
\addlegendentry{OQ}
\end{axis}
\end{tikzpicture}
\caption{Qwen1.5-7B-GPTQ-Int4}
\end{subfigure}

\caption{Effect of training steps under a fixed query budget (500 queries). CLIQ(ours) yields consistent gains over OQ and shows diminishing returns after $\sim$200--300 steps.}
\label{fig:steps_ablation}
\end{figure*}
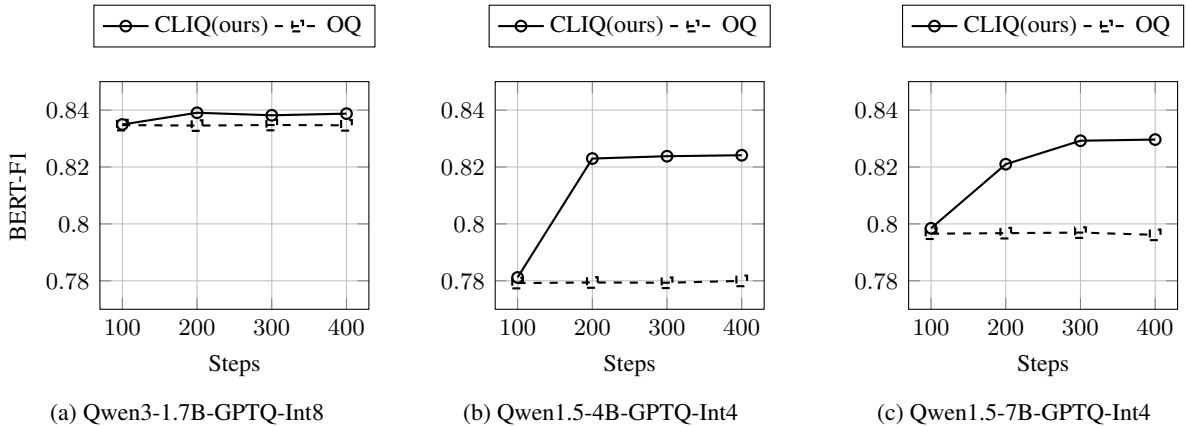


\paragraph{Different Quantization Settings.}
Table~\ref{tab:training_steps} further studies training steps under INT4
and INT8 quantization using the same 1.8B student and \emph{randomly sampled}
queries during training; the only difference is the construction of the query
pool (Original vs.\ CLIQ).
Two consistent patterns emerge across both quantization settings:
(i) with Original queries, performance remains nearly flat from 100 to 500 steps,
and can even slightly regress, and (ii) with CLIQ, performance improves sharply
from 100 to 300 steps and then saturates from 300 to 500 steps.
This contrast suggests that CLIQ provides a higher-quality query distribution
that the student can progressively absorb through optimization, whereas the
Original pool contains substantial redundancy that limits the benefit of
additional training.


\subsubsection{Clustering Granularity and Query Allocation}
\label{app:cluster_query_ablation}

Table~\ref{tab:dolly_and_alpaca} ablates two key design choices in CLIQ:
the number of semantic clusters and the number of generated queries per cluster,
evaluated on both Dolly and Alpaca.

\paragraph{Clustering Granularity.}
Across both datasets, moderate granularities (e.g., 25--50 clusters) tend to
produce stable and competitive results.
Overly fine-grained clustering (e.g., 100 clusters) exhibits higher variance and
can lead to performance degradation under certain allocations, consistent with
the intuition that excessive partitioning may fragment supervision into clusters
that are too small or semantically narrow.

\paragraph{Queries per Cluster and Diminishing Returns.}
Increasing queries per cluster generally improves performance when moving from 4
to 8 or 10 queries, but the gains diminish and are not strictly monotonic across
all settings.
This indicates that CLIQ benefits primarily from improved semantic coverage and
representative supervision rather than brute-force increases in query count.

\paragraph{Interaction Effects.}
We observe clear interactions between cluster count and per-cluster query
allocation.
Under finer clustering, insufficient queries per cluster can yield sparse and
unstable supervision, while allocating more queries partially mitigates this
issue.
Overall, these results suggest that balanced configurations (moderate clusters
with sufficient per-cluster queries) provide the best trade-off between semantic
coverage and query efficiency.

\paragraph{Theoretical Intuition and Empirical Alignment.}
The observed trends in Table~\ref{tab:dolly_and_alpaca} and
Figure~\ref{fig:query_budget} can be understood through a simple
coverage--redundancy trade-off.
Original query sampling corresponds to i.i.d.\ draws from a highly
imbalanced mixture of latent semantic clusters, where dense clusters
dominate early coverage while sparse clusters require substantially
larger budgets to be reliably sampled.
As a result, increasing the query budget under Original Queries often
leads to redundant supervision and early performance saturation.

In contrast, CLIQ enforces a structured allocation by distributing a
fixed number of queries across semantic clusters.
Under a fixed budget $B = K \cdot m$, increasing the number of clusters
$K$ improves global semantic coverage, while the number of queries per
cluster $m$ controls intra-cluster diversity.
This explains why moderate clustering granularities (e.g., $K=25$ or
$50$) combined with a small but sufficient number of queries per cluster
(e.g., $m=8$ or $10$) consistently achieve strong and stable performance
in Table~\ref{tab:dolly_and_alpaca}.
When $K$ is too large relative to the budget, insufficient per-cluster
queries lead to sparse and unstable supervision, whereas increasing $m$
beyond this range yields diminishing returns once dominant cluster modes
are captured.


\begin{table*}[t]
\centering
\small
\setlength{\tabcolsep}{6pt}
\renewcommand{\arraystretch}{1.15}
\begin{tabular}{l c c c c c c c c}
\toprule
\textbf{Dataset} & \textbf{Clusters} & \textbf{Queries} &
\textbf{BERT-F1} & \textbf{BLEU} &
\textbf{ROUGE-1} & \textbf{ROUGE-2} &
\textbf{ROUGE-L} & \textbf{ROUGE-Lsum} \\
\midrule
\multirow{7}{*}{Dolly}
 & 25  & 4  & 81.384 & 2.092 & 17.905 & 6.196 & 12.089 & 14.402 \\
 & 25  & 8  & 82.272 & 2.325 & 18.325 & 6.263 & 12.397 & 14.686 \\
 & 25  & 10 & 82.267 & 2.473 & 18.546 & 6.258 & 12.550 & 15.017 \\
 & 50  & 4  & 82.558 & 2.643 & 18.335 & 6.377 & 12.258 & 14.825 \\
 & 50  & 10 & 82.345 & 2.608 & 18.399 & 6.268 & 12.500 & 14.818 \\
 & 100 & 8  & 78.707 & 1.476 & 16.895 & 5.895 & 11.298 & 13.499 \\
 & 100 & 10 & \textbf{83.042} & \textbf{2.972} & \textbf{18.831} & \textbf{6.556} & \textbf{12.667} &\textbf{15.138} \\
\midrule
\multirow{7}{*}{Alpaca}
 & 25  & 4  & 81.582 & 1.861 & 17.788 & 5.711 & 11.650 & 14.146 \\
 & 25  & 8  & 81.451 & 2.010 & 17.749 & 5.873 & 11.563 & 14.039 \\
 & 25  & 10 & 81.936 & 2.256 & 18.177 & 6.088 & 11.977 & 14.508 \\
 & 50  & 4  & 80.914 & 1.885 & 17.701 & 6.060 & 11.669 & 14.165 \\
 & 50  & 10 & \textbf{82.518} & \textbf{2.460} &18.133 & 6.084 & 11.980 & 14.525 \\
 & 100 & 4  & 82.056 & 2.236 & \textbf{18.379} & \textbf{6.160} & \textbf{12.052} & \textbf{14.617} \\
 & 100 & 8  & 78.714 & 1.307 & 16.691 & 5.608 & 10.928 & 13.239 \\
\bottomrule
\end{tabular}
\caption{Additional ablation results on clustering granularity and query budget on a 4B model quantized by int4.
We vary the number of semantic clusters and the number of generated queries per cluster,
and report instruction-following performance on Dolly and Alpaca. All scores are reported in percentage (\%).}
\label{tab:dolly_and_alpaca}
\end{table*}



\subsection{Qualitative Analysis of Query Space}
\label{app:qualitative}
\subsubsection{Cluster-level Coverage and Redundancy Analysis}
\label{app:cluster_level_analysis}
\begin{figure*}
    \centering
    \arxivincludegraphics[width=\linewidth]{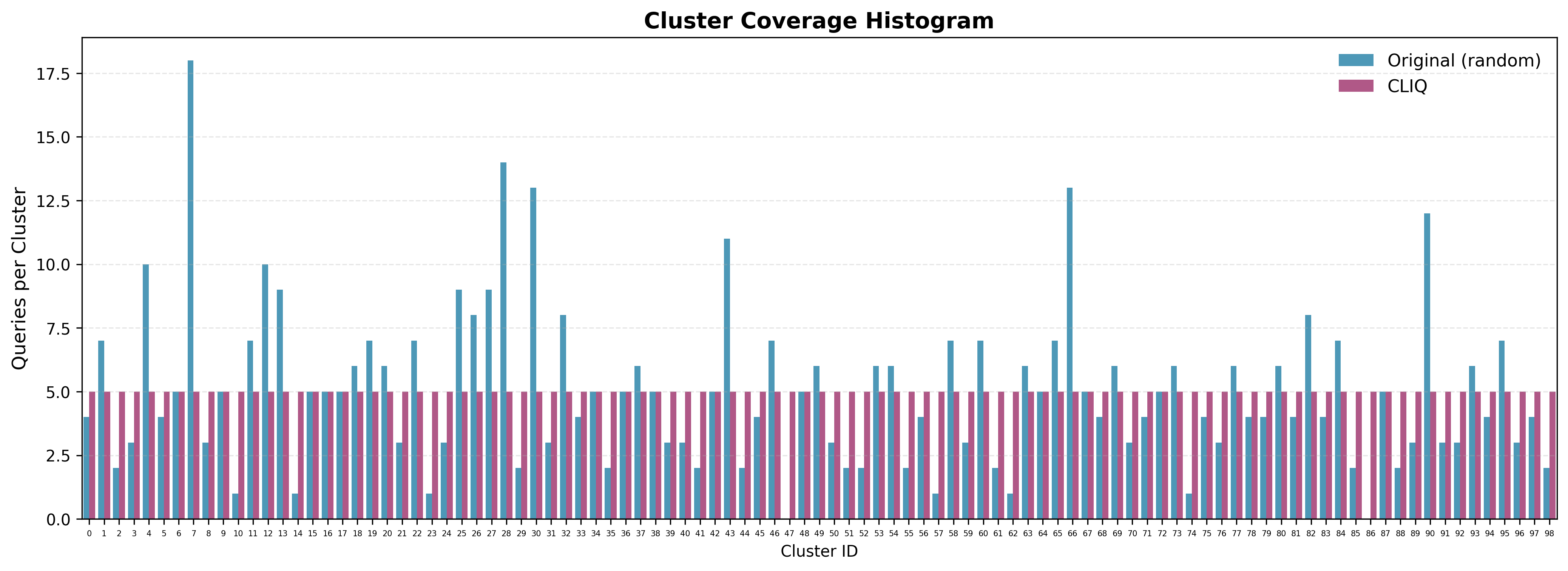}
    \caption{Histogram of query counts per cluster under a fixed query budget.
Random sampling from original queries results in highly imbalanced cluster
coverage, whereas CLIQ produces a near-uniform allocation across clusters,
avoiding over-sampling of dense regions.}
    \label{fig:cluster_histogram}
\end{figure*}

To better understand the mechanisms behind the quantitative gains observed in
Appendix~\ref{app:quant_results}, we further analyze how original queries (OQ)
and CLIQ-generated queries differ at the semantic cluster level.

\paragraph{Cluster Coverage under Fixed Query Budgets.}
\begin{figure}
    \centering
    \arxivincludegraphics[width=\linewidth]{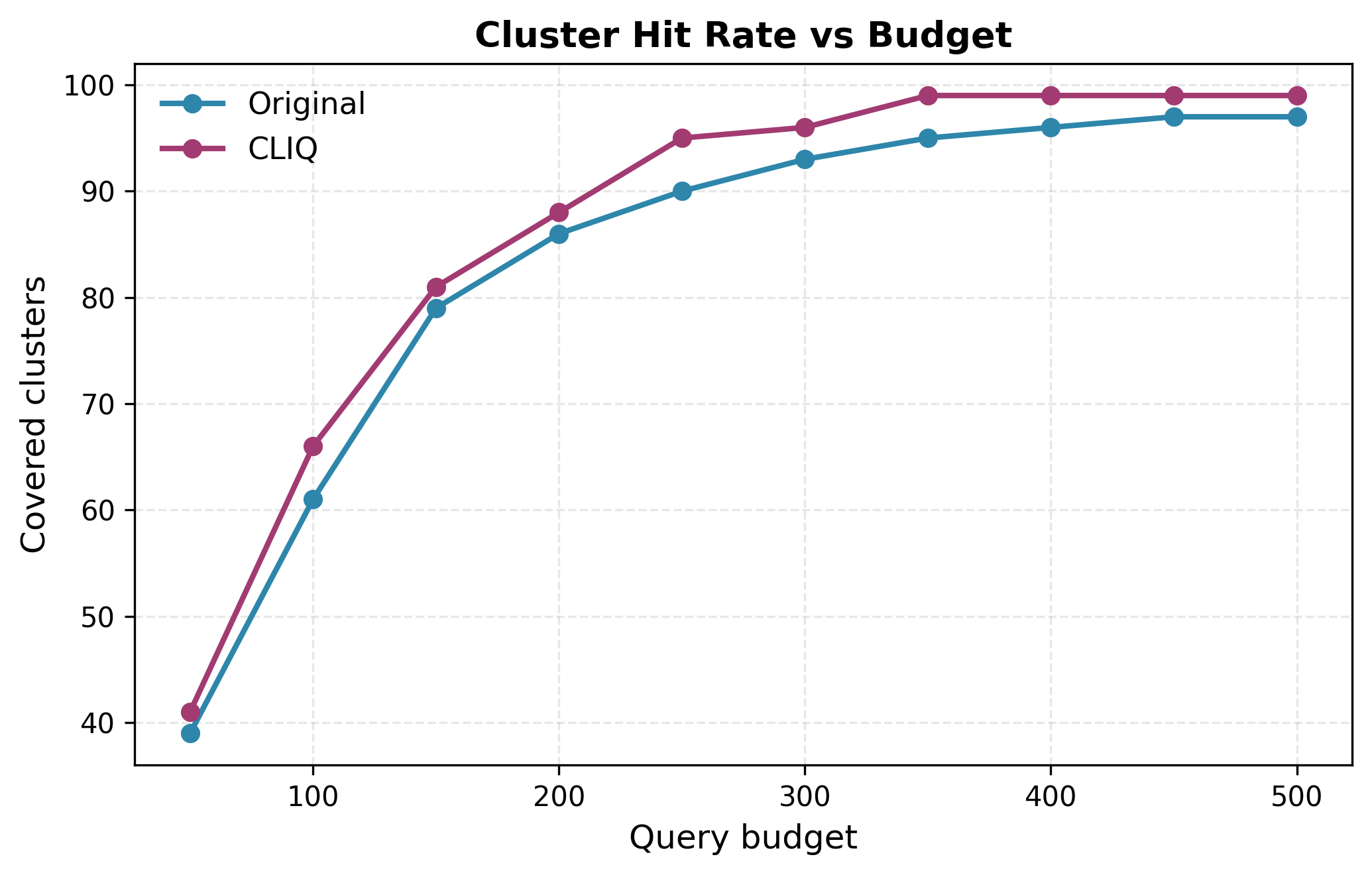}
    \caption{Cluster hit rate as a function of query budget.
We measure the number of semantic clusters covered by at least one query under
increasing query budgets.
Compared with random sampling from the original query pool (OQ), CLIQ achieves
substantially higher cluster coverage in the low-budget regime and reaches
near-complete coverage with fewer queries.
This indicates that CLIQ allocates queries more efficiently across semantic
clusters, particularly when the query budget is limited.}
    \label{fig:cluster_hit_rate}
\end{figure}

Figure~\ref{fig:cluster_hit_rate} reports the number of semantic clusters covered
as a function of the query budget.
While random sampling from the original query pool can eventually cover most
clusters as the budget increases, CLIQ consistently achieves higher coverage
under small to moderate budgets.
This early-stage advantage provides more informative supervision signals during
initial training, which helps explain the faster performance gains and earlier
saturation observed for CLIQ-trained models
(Section~\ref{app:training_dynamics}).

\paragraph{Intra-cluster Redundancy.}
\begin{figure}
    \centering
    \arxivincludegraphics[width=\linewidth]{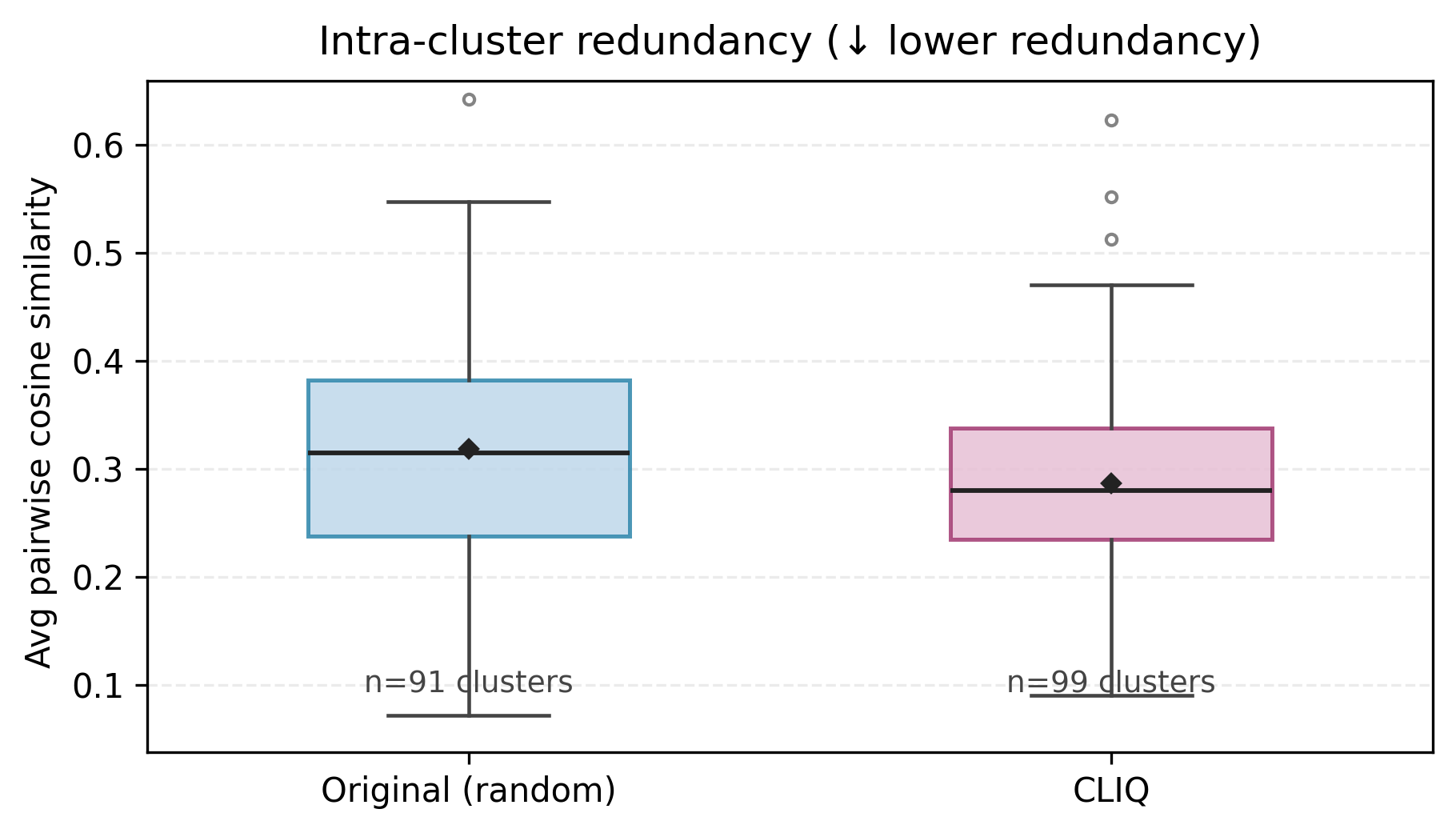}
    \caption{Intra-cluster redundancy measured by average pairwise cosine similarity within
each semantic cluster.
Original queries (OQ) exhibit significantly higher intra-cluster similarity,
indicating strong redundancy and overlapping supervision signals.
In contrast, CLIQ-generated queries reduce redundancy by promoting greater
diversity within clusters, which explains the improved sensitivity of
CLIQ-trained models to increased training steps and optimization.}
    \label{fig:intra_cluster_redundancy}
\end{figure}
Figure~\ref{fig:intra_cluster_redundancy} analyzes the average pairwise cosine
similarity among queries within each cluster.
Original queries exhibit substantially higher intra-cluster similarity,
indicating strong redundancy in supervision.
In contrast, CLIQ reduces redundancy by generating more diverse queries within
each cluster.
This observation explains why OQ-trained models are largely insensitive to
additional training steps, whereas CLIQ continues to yield gains as optimization
proceeds.

\paragraph{Distance to Cluster Centroids.}
\begin{figure}
    \centering
    \arxivincludegraphics[width=\linewidth]{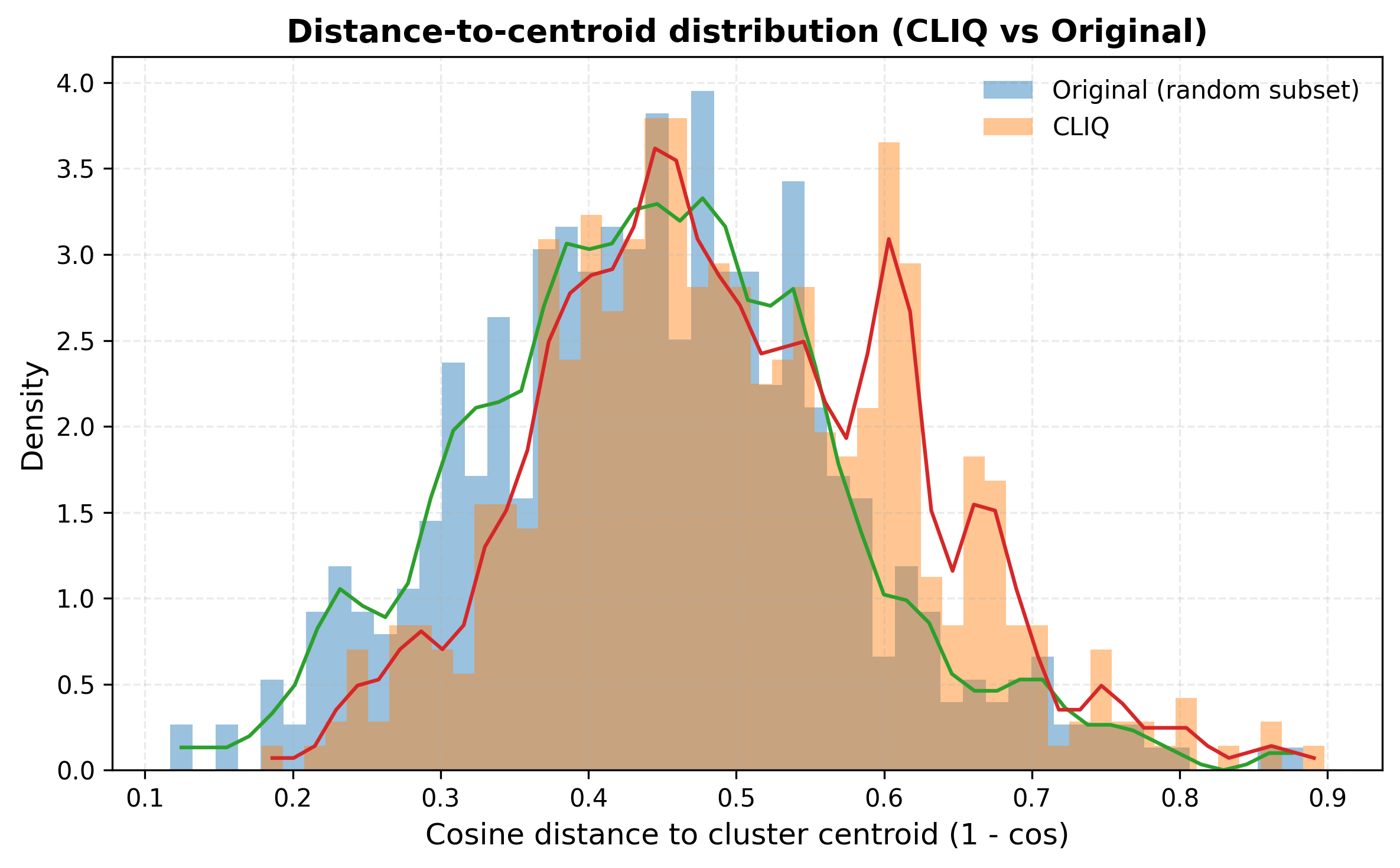}
    \caption{Distribution of cosine distances between queries and their corresponding cluster
centroids.
CLIQ-generated queries span a wider range of distances, covering both central and
peripheral regions of each cluster.
This indicates that CLIQ does not collapse queries to cluster prototypes, but
preserves structured intra-cluster diversity while maintaining semantic
coherence.}
    \label{fig:distance_to_centroid}
\end{figure}
Figure~\ref{fig:distance_to_centroid} shows the distribution of cosine distances
between queries and their corresponding cluster centroids.
CLIQ-generated queries span a wider range of distances, covering both central
and peripheral regions of each cluster.
This suggests that CLIQ does not collapse queries to cluster prototypes, but
instead preserves intra-cluster diversity while maintaining semantic coherence.

\paragraph{Cluster-level Query Allocation.}

Finally, Figure~\ref{fig:cluster_histogram} visualizes the number of queries
assigned to each cluster under a fixed query budget.
Random sampling from original queries results in highly imbalanced cluster
coverage, with certain dense clusters receiving disproportionate attention.
In contrast, CLIQ produces a near-uniform allocation across clusters, avoiding
over-sampling of dense regions and under-coverage of sparse ones.

Taken together, these analyses provide a unified explanation for the training
dynamics and ablation results reported in Appendix~\ref{app:quant_results}.
CLIQ improves performance not by increasing the total number of queries or
optimization steps, but by restructuring the query distribution to achieve
balanced cluster-level coverage, reduced redundancy, and greater semantic
diversity.

\subsubsection{Global Query Coverage}
\label{app:query_umap}

\begin{figure*}[t]
  \centering
  \arxivincludegraphics[width=\linewidth]{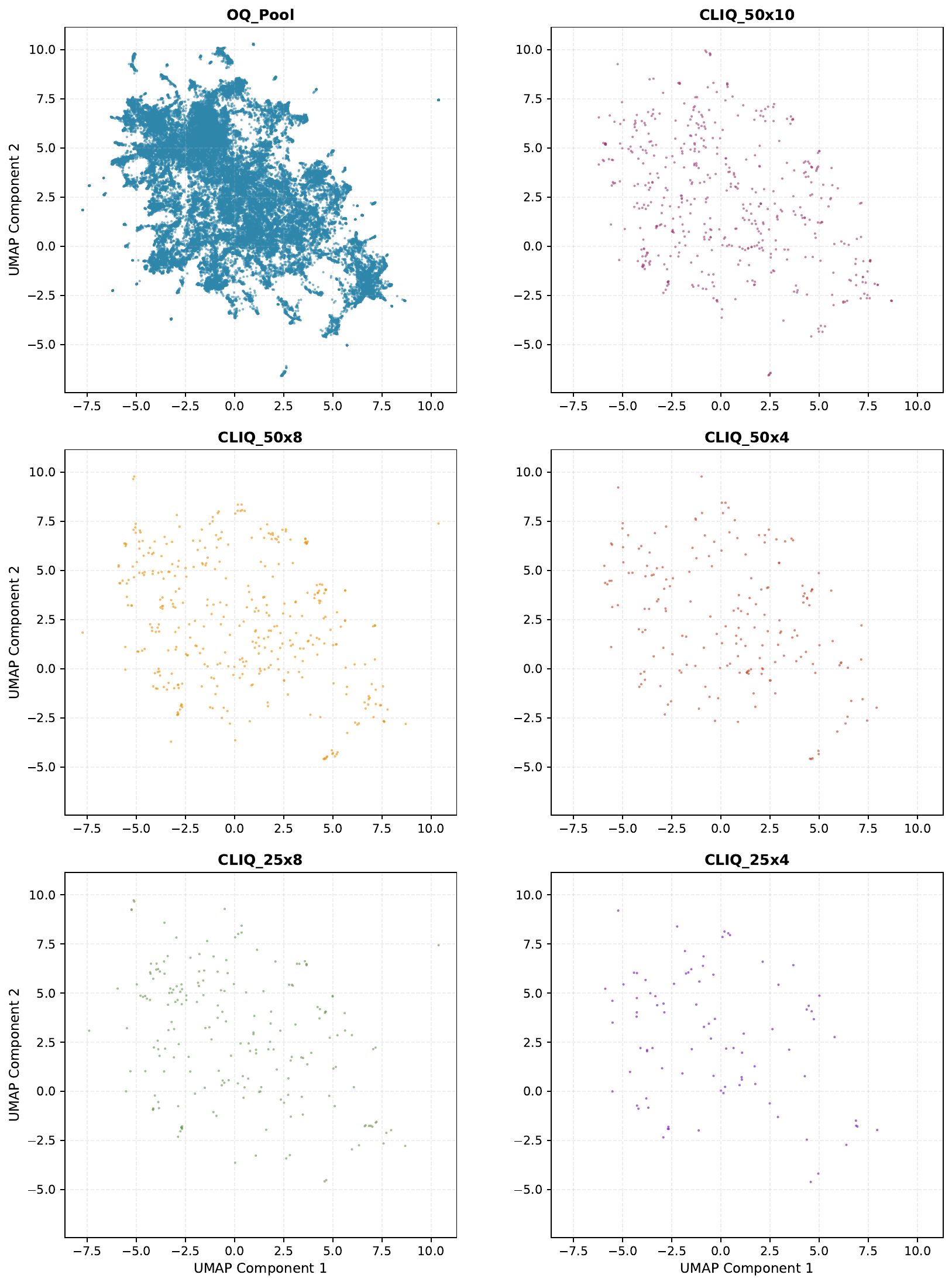}
  \caption{
    UMAP visualization of query coverage under different CLIQ configurations.
    The original query pool (OQ) exhibits dense and redundant regions, while
    CLIQ-generated queries maintain broader and more uniform semantic coverage
    under fixed query budgets. Different combinations of cluster numbers and
    queries per cluster illustrate how CLIQ controls coverage granularity.
  }
  \label{fig:UMP}
\end{figure*}

Figure~\ref{fig:UMP} visualizes the semantic distribution of different query pools
using UMAP \citep{UMAP} projections of sentence embeddings.
Each point corresponds to a single query, and all projections are computed using
the same embedding model for fair comparison.

\textbf{Original Query Pool.}
The original query pool (OQ) forms highly concentrated and dense regions in the
embedding space, with substantial overlap among queries.
This indicates strong semantic redundancy, where many queries occupy similar
regions and provide overlapping supervision signals.
Such redundancy offers a qualitative explanation for the limited sensitivity of
OQ-trained models to increased training steps and query budgets, as observed in
the training-dynamics and ablation results (Section~\ref{app:training_dynamics}).

\textbf{CLIQ-generated Query Pools.}
In contrast, query pools constructed by CLIQ exhibit significantly more dispersed
and uniform distributions.
Across different configurations, CLIQ expands semantic coverage while reducing
local redundancy, enabling a smaller number of queries to span a broader
instruction space.
This qualitative pattern is consistent with the quantitative gains reported in
Table~\ref{tab:dolly_and_alpaca}, where balanced CLIQ configurations achieve
strong performance with relatively modest query budgets.

\textbf{Effect of Clustering and Query Allocation.}
Varying the number of clusters and queries per cluster leads to clear and
interpretable changes in the query space structure.
Overly fine-grained clustering combined with insufficient queries per cluster
produces sparse coverage and visible gaps in the embedding space.
These gaps correspond to performance degradation in the quantitative ablation
(Table~\ref{tab:dolly_and_alpaca}), suggesting that excessive partitioning
fragments supervision into clusters that are too small to provide stable
learning signals.
Conversely, moderate clustering granularities with sufficient per-cluster queries
yield more uniform coverage and stronger downstream performance.

Taken together, the UMAP visualization provides a qualitative explanation for the
observed trade-offs between semantic coverage and query efficiency in CLIQ, and
supports the interpretation that performance gains arise from structured query
design rather than brute-force increases in query count.

\subsubsection{Cluster-conditioned Projections}
\label{app:cluster_projection}

\begin{figure*}[t]
  \centering
  \arxivincludegraphics[width=\textwidth]{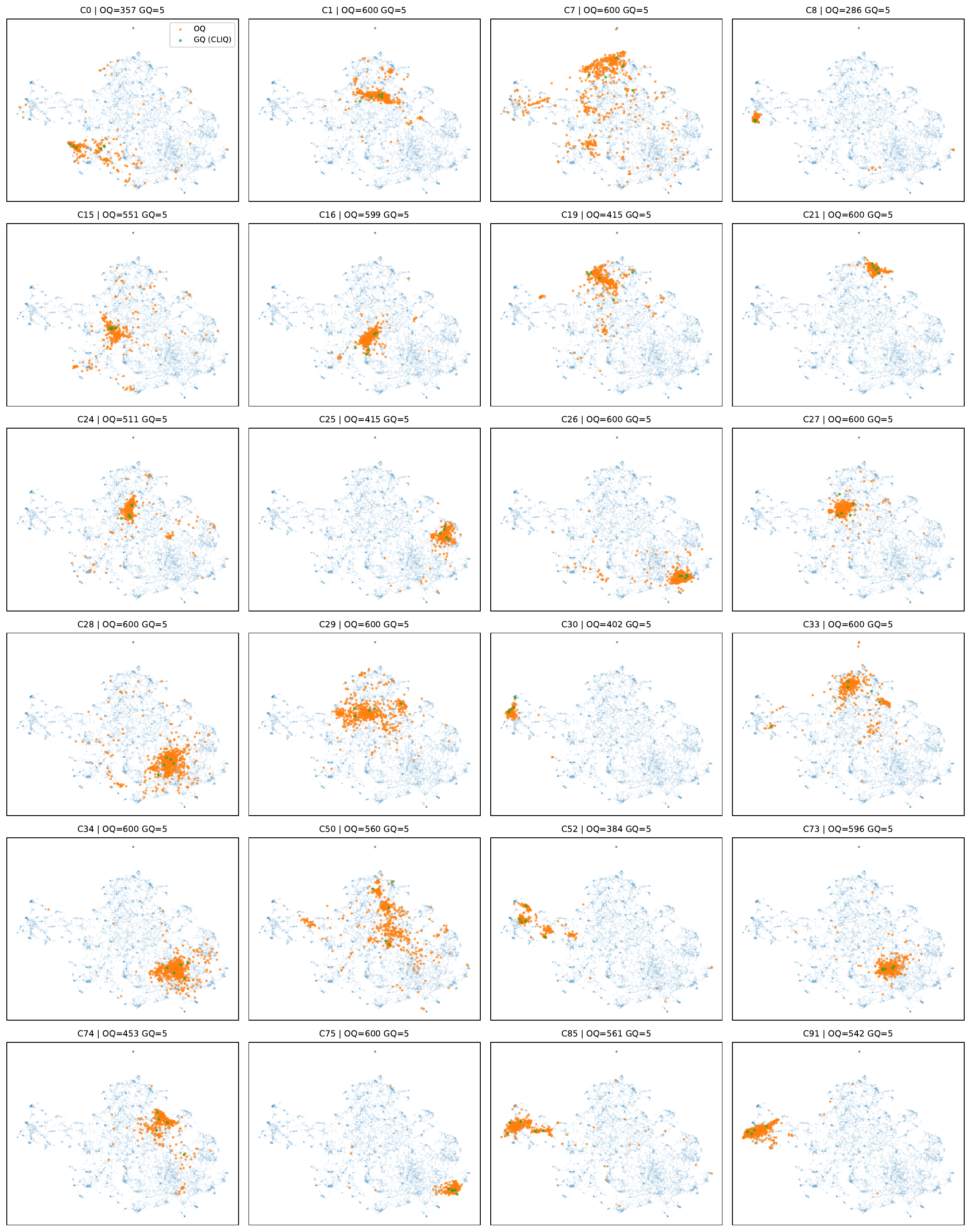}
  \caption{
    Cluster-conditioned semantic projections of original queries (OQ) and
    CLIQ-generated queries (GQ).
    Each subfigure corresponds to one semantic cluster $c$.
    Blue points denote original queries in the cluster, while orange points denote
    generated queries.
    Only $m=5$ generated queries per cluster are visualized to illustrate the
    query efficiency of CLIQ.
  }
  \label{fig:cluster_projection}
\end{figure*}

Figure~\ref{fig:cluster_projection} presents cluster-conditioned semantic
projections comparing original queries (OQ) and CLIQ-generated queries (GQ).
Each subfigure corresponds to one semantic cluster, where blue points denote
original queries and orange points denote generated queries.
Only $m=5$ generated queries per cluster are shown, highlighting the extreme
query-efficiency setting.

Despite the large imbalance between OQ and GQ, the generated queries consistently
align with high-density semantic regions of their corresponding clusters.
Rather than collapsing to global centroids, GQ samples often concentrate on
cluster-internal modes, indicating that cluster-conditioned prompting captures
salient semantic patterns specific to each cluster.

This observation complements the global UMAP analysis and the quantitative
cluster--query ablation results.
It provides qualitative evidence that CLIQ preserves semantic representativeness
within clusters while substantially reducing the number of required queries,
thereby explaining why CLIQ achieves strong performance under strict query
budgets in the main experiments.

\subsection{Use of AI Assistants}

AI-assisted tools were used in a limited and supportive manner during the preparation of this manuscript.
Specifically, large language models (e.g., ChatGPT) were employed to assist with language polishing,
clarity improvement, and structural refinement of the text, as well as minor LaTeX formatting suggestions.

The AI tools were not used to generate experimental results, derive theoretical claims, design algorithms,
conduct data analysis, or produce figures or tables.
All technical content, experimental design, analysis, and conclusions were conceived, verified,
and finalized by the authors.

The authors take full responsibility for the correctness, originality, and integrity of the content
presented in this paper.

\FloatBarrier


\end{document}